\RequirePackage{fix-cm}
\documentclass[twocolumn,epjc3]{svjour3}  

\RequirePackage{graphicx}
\usepackage{tikz}
\RequirePackage{mathptmx}
\usepackage[english]{babel}
\usepackage{amsmath}
\usepackage{mathtools}
\usepackage{siunitx}
\usepackage{comment}
\usepackage{tabularx}
\usepackage[version=4]{mhchem}
\usepackage[font={small},labelfont=bf,labelsep=quad]{caption}
\usepackage[font={normal}]{subcaption}
\usepackage{flushend}
\usepackage{cite}
\RequirePackage[colorlinks,citecolor=blue,urlcolor=blue,linkcolor=blue]{hyperref}

\defineshorthand{"~}{\babelhyphen{nobreak}}
\useshorthands{"}

\journalname{Eur. Phys. J. C}
\begin{document}
\sloppy

\title{A measurement of the mean electronic excitation energy of liquid xenon
}

\author{Laura Baudis
        \and
        Patricia Sanchez-Lucas
        \and
        Kevin Thieme\thanksref{e1}
}

\thankstext{e1}{e-mail: \href{mailto:kevin.thieme@physik.uzh.ch}{kevin.thieme@physik.uzh.ch}}

\institute{\normalsize{Department of Physics, University of Zurich, Winterthurerstrasse 190, 8057 Zurich, Switzerland}
}

\date{Received: date / Accepted: date}

\maketitle

\begin{abstract}
Detectors using liquid xenon as target are widely deployed in rare event searches. Conclusions on the interacting particle rely on a precise reconstruction of the deposited energy which requires calibrations of the energy scale of the detector by means of radioactive sources. However, a microscopic calibration, i.e.~the translation from the number of excitation quanta into deposited energy, also necessitates good knowledge of the energy required to produce single scintillation photons or ionisation electrons in liquid xenon. The sum of these excitation quanta is directly proportional to the deposited energy in the target. The proportionality constant is the mean excitation energy and is commonly known as $W$"~value. Here we present a measurement of the $W$"~value with electronic recoil interactions in a small dual-phase xenon time projection chamber with a hybrid (photomultiplier tube and silicon photomultipliers) photosensor configuration. Our result is based on calibrations at $\mathcal{O}(\SIrange{1}{10}{keV})$ with internal \ce{^{37}Ar} and \ce{^{83\text{m}}Kr} sources and single electron events. We obtain a value of $W=\SI{11.5}{} \, ^{+0.2}_{-0.3} \, \mathrm{(syst.)} \, \si{\electronvolt}$, with negligible statistical uncertainty, which is lower than previously measured at these energies. If further confirmed, our result will be relevant for modelling the absolute response of liquid xenon detectors to particle interactions.   
\end{abstract}

\section{Introduction}
\label{sec:intoduction}

Liquid xenon (LXe) is widely used as both sensitive target and radiation source in time projection chambers (TPCs) for present and future rare event searches. These include e.g.~searches for neutrinoless double beta decay~\cite{EXO-200:2019rkq,nEXO:2017nam,Agostini:2020adk}, the extremely rare decay of \ce{^{124}Xe} via double electron capture~\cite{XENON:2019dti,LUX:2019zsx}, the measurement of low-energy solar neutrinos~\cite{Aalbers:2020gsn}, searches for solar axions, axion-like particles and dark photons~\cite{Aprile:2020tmw,LUX:2017glr,LZ:2021xov} as well as for WIMP (weakly interacting massive particle) dark matter in the $\si{GeV}$ to $\si{TeV}$ mass range. In particular, TPCs operated in dual-phase mode with a gaseous xenon (GXe) layer at the top are among the leading technologies in the past, present and near-future hunt for WIMPs~\cite{Aprile:2018dbl,Akerib:2016vxi,Cui:2017nnn,Aprile:2020vtw,LUX-ZEPLIN:2018poe,Zhang:2018xdp,Aalbers:2016jon}.   
 
A particle that deposits energy in LXe yields scintillation photons in the vacuum-ultraviolet (VUV) range with a peak centred at $\SIrange{175}{178}{nm}$ wavelength~\cite{Jortner:1965,FUJII2015293}, ionisation electrons and heat via atomic motion, where only the former two processes are detectable with dual-phase TPCs. Photosensors detect the prompt scintillation light ($S1$) which is composed of direct scintillation of the xenon atoms and light from the recombination of xenon ions with electrons. In both mechanisms, scintillation light is produced by de-excitation of xenon dimers that form from xenon excitons and xenon atoms. Electrons that do not recombine are vertically drifted and extracted to the gas phase, by means of an electric drift and extraction field, where they collide with xenon atoms. In this process, a secondary proportional scintillation signal ($S2$) is produced by electroluminescence which is also detected by photosensors. Due to energy conservation, the size of the signals from scintillation and ionisation are anti-correlated~\cite{EXO-200:2003bso}. The recombination fraction and thus the energy distribution between these is drift field dependent. We collectively refer to scintillation and ionisation as excitation in this work.

The number of scintillation photons $n_{\gamma}$ and ionisation electrons $n_{\mathrm{e}^{-}}$ in an electronic recoil interaction is linearly related to the deposited energy $E$ by the constant work function $W$~\cite{Dahl:2009nta}:
\begin{equation}
\label{eq:W_value_def}
E=(n_{\gamma}+n_{\mathrm{e}^{-}})W \quad.
\end{equation}
The $W$"~value can be regarded as the average energy needed to produce a single free quantum in LXe and the above expression as its defining equation. It determines the underlying recombination-independent energy scale in a LXe detector that detects both scintillation light and ionisation charge. Note that we implicitly assume here that every recombining electron-ion pair leads to the production of one photon. 

The widely employed numerical value of $W$ is $(\SI{13.7}{} \pm \SI{0.2}{}) \, \si{eV}$ measured at $\sim \SI{100}{keV}$ by E.~Dahl~\cite{Dahl:2009nta}. However, a $W$"~value of $(11.5 \pm 0.1\,\mathrm{(stat.)} \pm 0.5\, \mathrm{(syst.)}) \, \si{eV}$ was measured in the EXO"~200 experiment with various gamma sources at $\mathcal{O}(\SI{1}{MeV})$~\cite{EXO-200:2019bbx}. A list of other measurements of the $W$"~value can be found in the same reference. The deviation of the EXO"~200 value from former measurements motivated this study at keV-scale energies, deploying internal \ce{^{37}Ar} and \ce{^{83\text{m}}Kr} sources. The measurements were performed in our small dual-phase xenon TPC Xurich~II, which is equipped with a bottom photomultiplier tube (PMT) and a silicon photomultiplier (SiPM) top array~\cite{Baudis:2020nwe}. 

This article is structured as follows: In Sec.~\ref{sec:experimental_approach} we outline the deployed experimental method, introduce briefly the setup and the data on which we base our analysis. In Sec.~\ref{sec:measurements} we show the data analysis of the various inputs for the calculation of the $W$"~value. The systematic uncertainties on the measurement and the applied data corrections are detailed in Sec.~\ref{sec:systematics_corrections}. In particular, we provide a study of a possible double photoelectron emission in SiPMs in~\ref{app:SiPM_DPE}. We present the final result in Sec.~\ref{sec:result} and discuss, interpret and summarise our findings in Sec.~\ref{sec:discussion_conclusion}. 
\section{Experimental approach}
\label{sec:experimental_approach}

\subsection{Measurement principle}
\label{subsec:method}

\begin{figure}
\centering
\begin{tikzpicture}
\node [black] at (5.5,5.5) {\normalsize{\textit{Doke plot}}};
\draw [black, very thick, ->] (0,0) -- (6,0) node[right] {\normalsize{LY}};
\draw [black, very thick, ->] (0,0) -- (0,6) node[above] {\normalsize{QY}};
\draw [black, thick] (0,2) -- (2,0);
\draw [black, thick, dashed] (0,5) -- (5,0);
\draw [black] (-0.1,2) -- (0.1,2);
\draw [black] (2,-0.1) -- (2,0.1);
\draw [black] (-0.1,5) -- (0.1,5);
\draw [black] (5,-0.1) -- (5,0.1);
\node [black, anchor=east] at (0,2) {\normalsize{$\left. \frac{S2}{E} \right|_{S1=0}$}};
\node [black, anchor=north] at (2,0) {\normalsize{$\left. \frac{S1}{E} \right|_{S2=0}$}};
\draw [black, thick, ->] (-0.2,3) -- (-0.2,4) node[left] {\normalsize{$\times W$}};
\draw [black, thick, ->] (3,-0.2) -- (4,-0.2) node[below] {\normalsize{$\times W$}};
\node [black, anchor=east] at (0,5) {\normalsize{$g2$}};
\node [black, anchor=north] at (5,0) {\normalsize{$g1$}};
\draw [black] (3,1) -- (4,1);
\draw [black] (3,1) -- (3,2);
\node [black, anchor=east] at (3,1.5) {\normalsize{$\propto -g2$}};
\node [black, anchor=north] at (3.5,1) {\normalsize{$\propto g1$}};
\node [black] at (4,1.6) {\normalsize{$\propto -\frac{g2}{g1}$}};
\draw [black] (0.9,4) -- (1.1,4);
\draw [black] (1,3.9) -- (1,4.1);
\node [black, anchor=west] at (1.1,4) {\normalsize{$S2\frac{W}{E}=g2-\frac{g2}{g1}S1\frac{W}{E}$}};
\draw [black] (0.3,1.6) -- (0.5,1.6);
\draw [black] (0.4,1.5) -- (0.4,1.7);
\node [black, anchor=west] at (0.5,1.6) {\normalsize{Data}};
\end{tikzpicture}
\caption{Schematic of the \textit{Doke plot} in charge yield (QY) versus light yield (LY) space for an interaction energy $E$. Measurements of the charge and light yield at different electric drift fields or interaction energies would yield the solid line. The dashed line, making the relation to the gain parameters $g1$ and $g2$ apparent, is obtained by a scaling of this space with $W$. While the local approach uses a point on the anti-correlation line and its (negative) slope $g2/g1$, the global one uses the (reciprocal) QY-axis intercept. Both require an independent measurement of the gain parameter $g2$.}
\label{fig:Doke_schematic}
\end{figure}
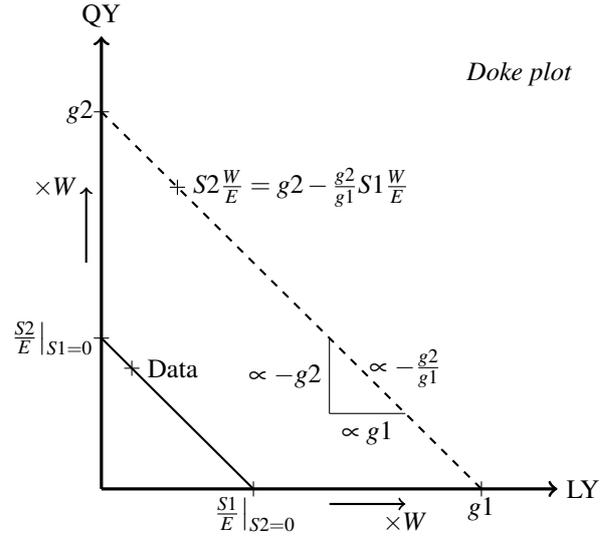

We rewrite Eq.~\ref{eq:W_value_def} with the scintillation gain $g1 \coloneqq S1/n_{\gamma}$ and the ionisation gain $g2 \coloneqq S2/n_{\mathrm{e}^{-}}$, such that $W$ reads
\begin{equation}
\label{eq:W_value_det1}
W=g2\frac{E}{\frac{g2}{g1} S1+S2} \quad.
\end{equation}
Thus, the $W$"~value can be determined from the following inputs: an event population in $S1$"~$S2$"~space of a known calibration source yielding electronic recoils at an energy $E$, an independent measurement of the gain parameter $g2$ and the (negative) slope $g2/g1$ of the first order polynomial in ionisation (charge) yield versus scintillation (light) yield space. We refer to the latter representation of the anti-correlation of the $S1$~and $S2$~signals as \textit{Doke plot}~\cite{Doke:2002} and show a schematic in Fig.~\ref{fig:Doke_schematic}.

Alternatively to this local approach, evaluated at a certain energy and scaled by the slope of the anti-correlation line in the \textit{Doke plot}, we can extrapolate to $S1=0$, i.e.~to the intercept with the charge yield axis, and obtain the global expression (see Fig.~\ref{fig:Doke_schematic})
\begin{equation}
\label{eq:W_value_det2}
W=\left. g2\frac{E}{S2} \right|_{S1=0} \quad.
\end{equation}
Besides the gain parameter $g2$, the only input here is the extrapolated (reciprocal) offset of the anti-correlation line at zero light yield. Both the ratio of gains $g2/g1$ as well as the offset $S2/E$ at $S1=0$ require at least $S1$~and $S2$~data of a single energy line at two different electric drift fields or two different energy lines at one electric drift field. For more than two light and charge yield pairs, one performs a linear fit in the \textit{Doke plot} whose accuracy depends on the separation in the $S2/S1$~ratio and on the yield errors.  

Besides the parameters of the anti-correlation fit, both approaches require an independent measurement of the gain parameter $g2$ as input. To that end, we observe that for a single electron (SE) that is extracted to the gas phase, we have $g2=S2$. Hence, a measurement of the SE~event population with the $S2$~signal yields the gain parameter $g2$.   

\subsection{Experimental setup}
\label{subsec:setup}

\begin{figure*}
\centering
\begin{subfigure}[b]{0.49\textwidth}
\includegraphics[width= \textwidth]{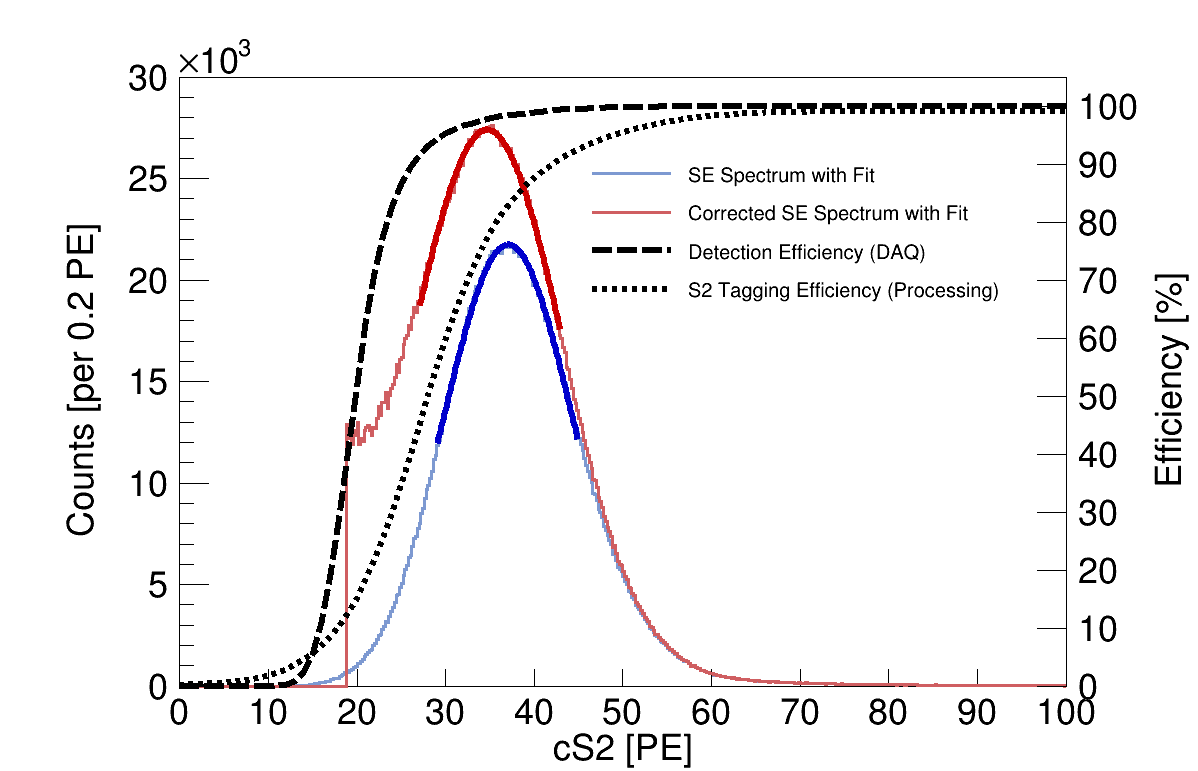}
\end{subfigure}
\begin{subfigure}[b]{0.49\textwidth}
\includegraphics[width= \textwidth]{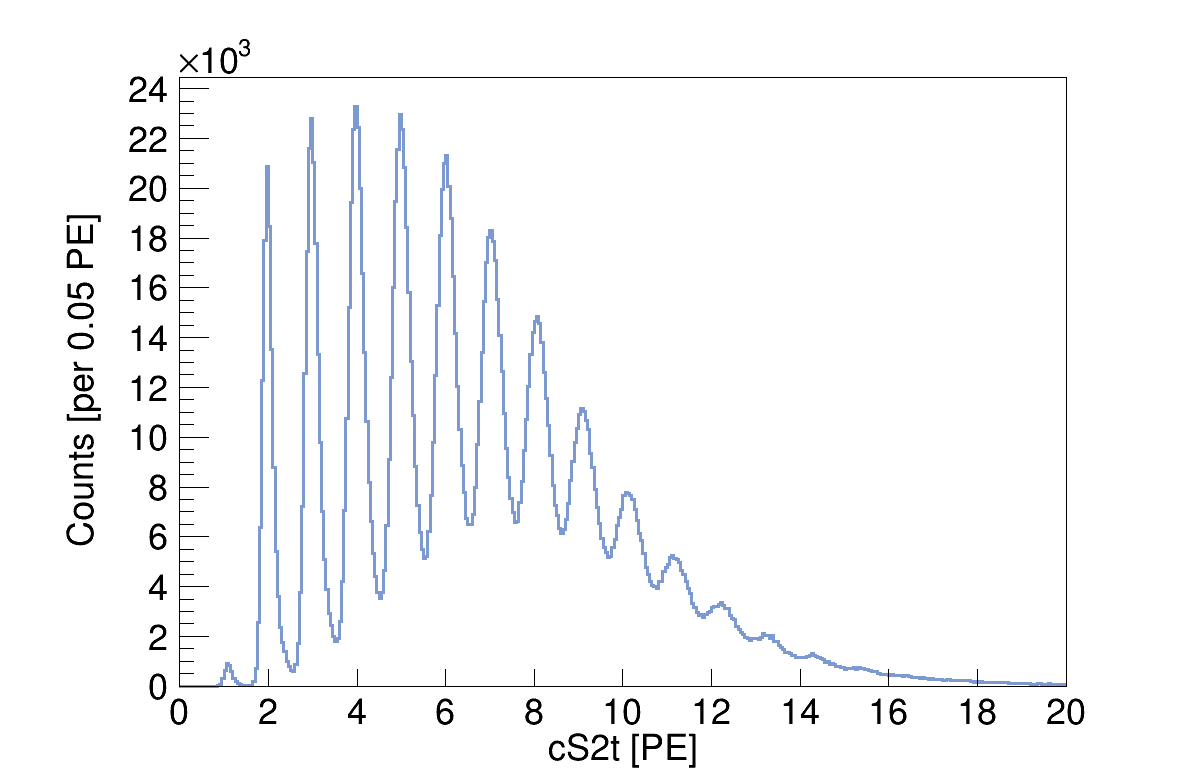}
\end{subfigure}
\caption{Total $S2$~population from SE~extraction at $\SI{968}{V/cm}$ drift field without DPE/crosstalk correction. Left: Bare and efficiency-corrected $S2$~total spectrum with Gaussian fits together with the efficiency curves, see Sec.~\ref{subsubsec:detection_tagging_efficiency}. The corrected (red) spectrum was cut on the left to avoid division by small efficiencies ($< \SI{5}{\%}$) that distort the distribution. Right: Bare $S2$~top ($S2$t) spectrum with single PE resolution. The PE"~scale of the $S2$t~signals is fixed by the centring of the peaks around the corresponding PE"~tick mark using a linear correction function. We follow the same procedure for the $S1$t but base it on the $\SI{2.82}{keV}$ population of \ce{^{37}Ar}, cf.~Ref.~\cite{Baudis:2020nwe}, Sec.~5.1.2.}
\label{fig:se_population}
\end{figure*}

Xurich~II is a dual-phase xenon TPC with a cylindrical $\SI{31}{mm} \times \SI{31}{mm}$ drift region that provides external electric drift fields of up to $\sim \SI{1}{kV/cm}$, applied between the cathode and the gate mesh. For the presented data, it was operated with a $\SI{10}{kV/cm}$ ($\SI{5.4}{kV/cm}$) gas (liquid) extraction field, applied between the gate and the anode mesh which are at $\SI{4}{mm}$ distance. It is equipped with a single 2"~inch PMT (R6041"~06~MOD, Hamamatsu Photonics) at the bottom and a $4\times4$ SiPM array (2x2 array of S13371, Hamamatsu Photonics) at the top and allows for a precise three-dimensional event position reconstruction with a resolution of $\sim \SI{1.5}{mm}$ in the horizontal plane. The detector was operated such that the GXe was kept at a pressure of $\SI{2.0}{bar}$ and the LXe was slightly below its boiling point at $\SI{177}{K}$. Under these conditions, LXe has a density of $\SI{2861}{kg/m^3}$~\cite{NIST}. The detector was kept in this state during the calibration runs and the thermodynamical parameters only varied slightly over the data acquisition periods: the pressure and LXe temperature range was $\SI{0.05}{bar}$ and $\SI{0.4}{K}$, respectively. The gas recirculation flow through the hot metal getter, crucial for the LXe purity (see Sec.~\ref{subsubsec:drift_time}), was stable within $\SIrange{0.5}{0.6}{slpm}$ (slpm -- standard litre per minute). A detailed description of the device, the data acquisition (DAQ) and processing, the event reconstruction as well as the \ce{^{37}Ar} and \ce{^{83\text{m}}Kr} analysis procedures can be found in Ref.~\cite{Baudis:2020nwe}.

\subsection{Data}
\label{subsec:data}

We base our analysis on high-statistics calibration data in $S1$"~$S2$"~space from two different runs with internal \ce{^{37}Ar} and \ce{^{83\text{m}}Kr} sources~\cite{Baudis:2020nwe}. \ce{^{37}Ar} provides an energy line at $\SI{2.82}{keV}$ from a K"~shell electron capture~\cite{Barsanov:2007fe}. \ce{^{83\text{m}}Kr} offers two energy lines at $\SI{32.15}{keV}$ and $\SI{9.41}{keV}$ from an isometric transition. The two lines are the result of an intermediate decay state with a half-life of $\SI{155.1}{ns}$~\cite{McCutchan:2015}. If they are not resolved separately at short delay times, due to detection or data processing limitations, or if they are purposely merged in the data analysis, we observe the combined energy of $\SI{41.56}{keV}$. Data sets acquired at electric drift fields of $\SIrange{80}{968}{V/cm}$ for \ce{^{37}Ar} and of $\SIrange{484}{968}{V/cm}$ for \ce{^{83\text{m}}Kr} are used.

As mentioned in Sec.~\ref{subsec:method}, the ionisation gain parameter $g2$ is determined based on the observed SE~events with $S2$~signals only. These events occur at a high rate of $\sim \SI{17}{Hz}$ in the detector volume and are homogeneously distributed in the horizontal plane. However, using a centre-of-gravity algorithm as described in Ref.~\cite{Baudis:2020nwe}, it is clear that SE~events with e.g.~only one photon in the top array are reconstructed at the centre of the hit photosensor. This natural limitation of the position reconstruction in the single photon regime does however not affect the analysis. A systematical variation of the fiducial radius did not show any significant change of the bottom, top or total SE~charge yield up to radii close to the TPC boundary. The SE~event population in dual-phase TPCs is well-known but its origin not yet fully understood~\cite{XENON100:2013wdu,Edwards:2007nj,ZEPLIN-III:2011qer,Akimov:2016rbs,Kopec:2021ccm}. In the given references, a time-correlation to high-energy depositions is observed as well as a dependence on the concentration of electronegative impurities in the LXe. Single electrons can originate from a delayed extraction to the gas phase, from trapped charge on the TPC surfaces or from electron emission at the cathode. Other production mechanisms involve stimulation by the VUV scintillation light, such as photodetachment from impurities or photoelectric effect on metal surfaces~\cite{XENON100:2013wdu}. In case of the deployed \ce{^{37}Ar} source, the M"~shell electron capture process at $\SI{17.5}{eV}$~\cite{Barsanov:2007fe} could cause, based on the branching ratio with the K"~shell, a small fraction of $\sim \SI{0.5}{\%}$ of this SE~population. The origin of these electrons is however not relevant for this analysis as they all share the same signal topology. We extract the SE~population from the \ce{^{37}Ar} data sets since these were acquired with a low trigger threshold of $\SI{7}{mV}$ on the PMT channel~\cite{Baudis:2020nwe}. Because of the highly abundant one by one extraction of electrons at high $g2$\footnote{Based on our previous analysis~\cite{Baudis:2020nwe}, we expect a high $g2$ of $\SIrange{30}{40}{PE/e^{-}}$ (PE -- photoelectron) for an assumed $W$"~value of $\SIrange{11}{14}{eV}$.}, the SE~population is readily identified and isolated (cf.~Fig.~\ref{fig:se_population}). 
\section{Data analysis}
\label{sec:measurements}

\subsection{Single electron gain}
\label{subsec:single_electron_gain}

\begin{figure*}
\centering
\begin{subfigure}[b]{0.49\textwidth}
\includegraphics[width= \textwidth]{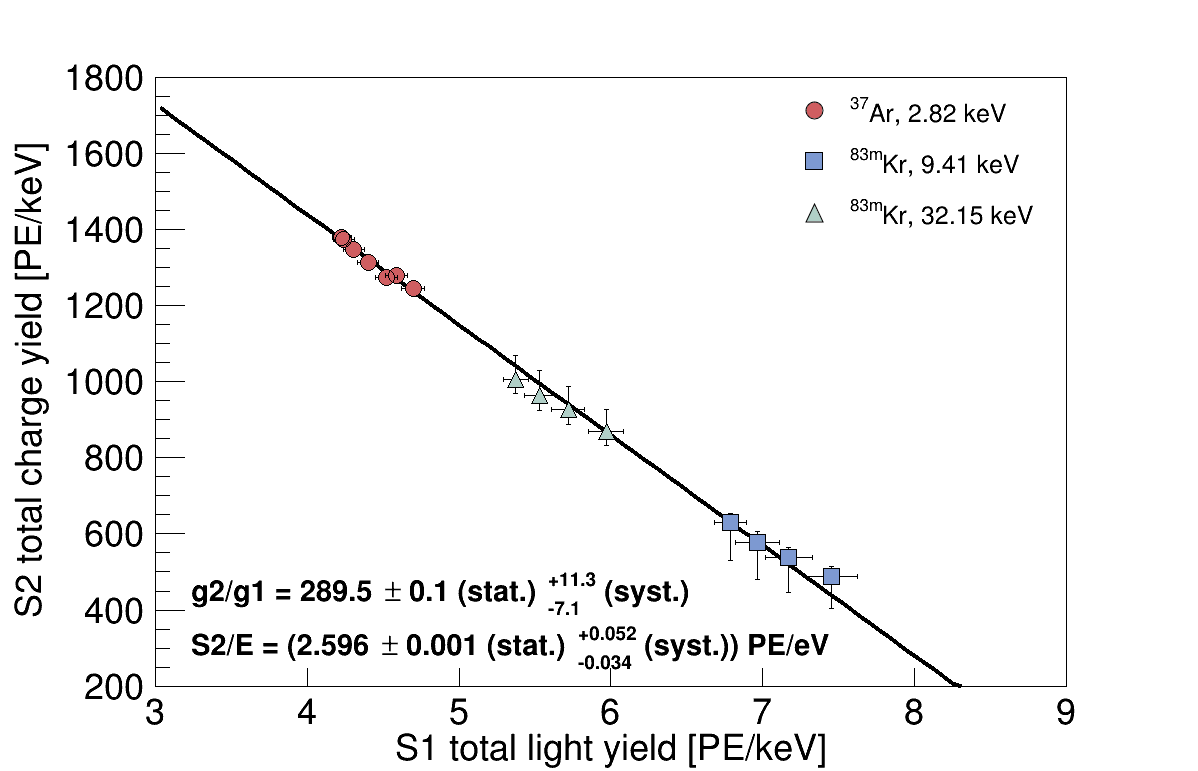}
\end{subfigure}
\begin{subfigure}[b]{0.49\textwidth}
\includegraphics[width= \textwidth]{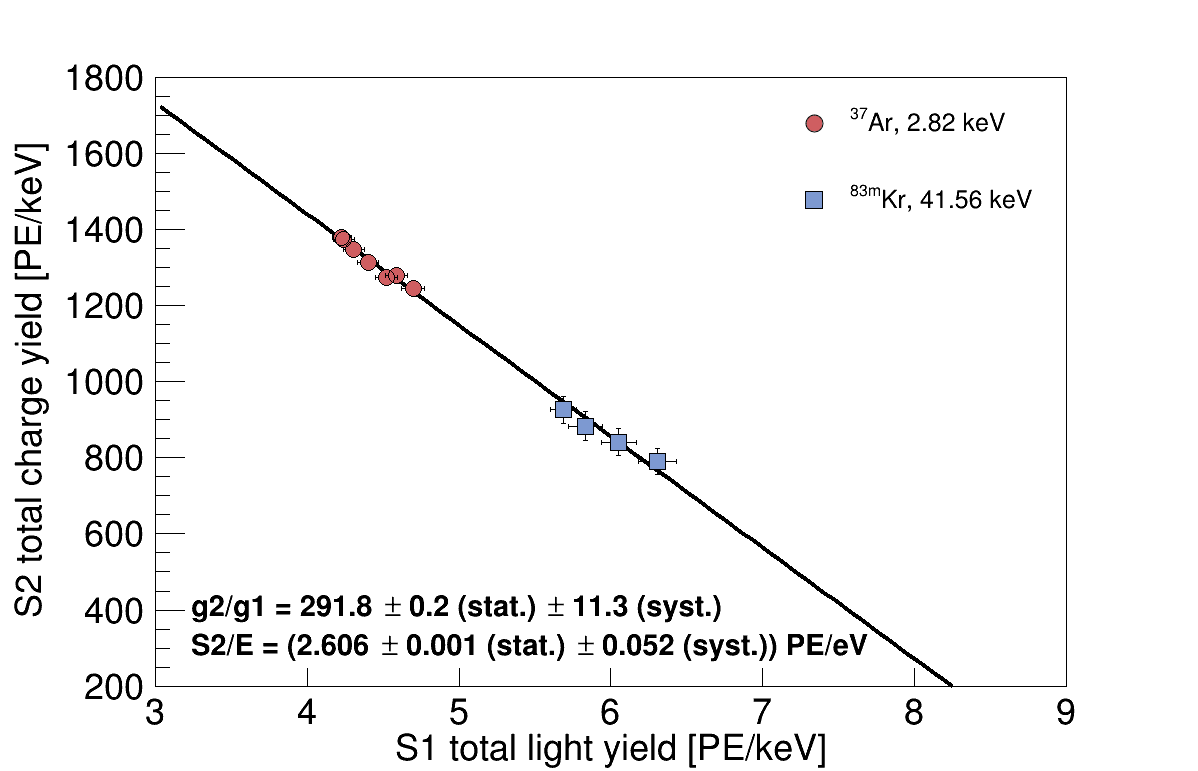}
\end{subfigure}
\caption{Anti-correlated charge versus light yield of \ce{^{37}Ar} and \ce{^{83\text{m}}Kr} at various drift fields with linear fit. The error bars are updated compared to Ref.~\cite{Baudis:2020nwe}. Left: Including the split $\SI{9.41}{keV}$ and $\SI{32.15}{keV}$ \ce{^{83\text{m}}Kr} lines with asymmetric error bars from the splitting routine, see Sec.~\ref{subsubsec:peak-splitting_routine}. Right: Including the merged $\SI{41.56}{keV}$ line.}
\label{fig:doke_plot}
\end{figure*}

To identify and isolate the SE~event population, we apply the following data selection and quality criteria: a cut to remove saturated events, a single scatter cut on the PMT channel and a fiducial radius cut such as developed in Ref.~\cite{Baudis:2020nwe}. In addition, we apply an $S2$"~only cut, i.e.~we remove events with an $S1$~signal in the PMT waveform. Furthermore, we require that the trigger was issued on the $S2$~signal, i.e.~we keep events whose PMT $S2$~signal is located within a tight time window of $[-100,+130]\, \si{ns}$ around the trigger. An $S2$"~width cut of $[110,550]\, \si{ns}$ further cleans up the population. The $S2$"~width increases with the depth of the interaction in the TPC due to the vertical diffusion of the electron cloud during the drift. Higher electric fields allow for higher drift speeds that reduce the effect of diffusion~\cite{Sorensen:2011qs}. 

The mean of the Gaussian fit of the population is constant within the parameter uncertainty for the considered drift fields. We show the result for the $\SI{968}{V/cm}$ data sets in Fig.~\ref{fig:se_population}. A Gaussian fit of the bare, i.e.~non-corrected, spectrum yields $g2=(\SI{37.05}{} \pm \SI{0.01}{} \, (\mathrm{stat.}) \pm 0.50 \, (\mathrm{syst.})) \, \si{PE/e^{-}}$. Incorporating the DPE/crosstalk correction as well as the DAQ and processing efficiencies, discussed in Secs.~\ref{subsubsec:DPE_crosstalk} and~\ref{subsubsec:detection_tagging_efficiency}, we obtain $g2=(\SI{29.84}{} \pm \SI{0.01}{} \, (\mathrm{stat.}) \pm \SI{0.40}{} \, (\mathrm{syst.})) \, \si{PE/e^{-}}$. While the DPE/crosstalk correction shifts the SE~spectrum by $\sim \SI{14}{\%}$, the DAQ and processing efficiencies contribute by another $\sim \SI{6}{\%}$.

\subsection{Anti-correlation fit parameters}
\label{subsec:doke_fit_parameters}

In Fig.~\ref{fig:doke_plot}, we show the means of the \ce{^{37}Ar} and \ce{^{83\text{m}}Kr} populations at the considered drift fields in charge versus light yield space (see Ref.~\cite{Baudis:2020nwe} for the elliptical fits of the populations in $S1$"~$S2$"~space and also Fig.~\ref{fig:s2_delay_K-shell_population}, right). In the left plot, we show the fit with the split $\SI{32.15}{keV}$ and $\SI{9.41}{keV}$ lines and in the right one the merged $\SI{41.56}{keV}$ line. Although the $\SI{41.56}{keV}$ line is not subject to the effect of the peak-splitting routine, that is discussed in Sec.~\ref{subsubsec:peak-splitting_routine}, it does not yield more precise results of the fit parameters on the anti-correlation line due to its smaller span in $S1$"~$S2$"~space. However, we find both fits to be compatible within errors. From the linear fit on the left of Fig.~\ref{fig:doke_plot} we obtain a (negative) slope of $g2/g1=289.5 \pm 0.1 \, (\mathrm{stat.}) \, ^{+11.3}_{-7.1} \, (\mathrm{syst.})$ and a charge yield axis intercept of $S2/E=(\SI{2.596}{}\pm \SI{0.001}{} \, (\mathrm{stat.}) \, ^{+0.052}_{-0.034} \, (\mathrm{syst.})) \, \si{PE/\electronvolt}$.
\section{Systematic uncertainties and corrections}
\label{sec:systematics_corrections}

A determination of the absolute energy scale of a LXe detector requires a careful treatment of the systematic uncertainties and efficiencies that impact the result. Below, we discuss in detail both relevant sources and those that were identified to be negligible, as well as data corrections that were applied whenever possible. We provide a summary of all treated systematic effects and corrections in Table~\ref{tab:systematics}.  

\subsection{TPC effects}
\label{subsec:TPC_effects}

\subsubsection{Liquid xenon purity and TPC geometry}
\label{subsubsec:drift_time}

For a given energy deposition, the size of the $S1$~and $S2$~signals depend on the vertical depth of the interaction in the TPC and thus, on the drift time. Electrons can attach to electronegative impurities in the LXe during their drift. This charge loss reduces the size of the $S2$~signal with increasing drift time and is quantified by the free electron lifetime. In addition, the light collection in the top and bottom photosensors is subject to the geometry of the TPC, and to the optical properties of the TPC surfaces and the LXe, and thus also features a dependence on the drift time. To minimise geometry effects, ensure good performance of the position reconstruction algorithm, and to reduce the material-induced background and the effect of electric field distortions, we fiducialise our active volume like in Ref.~\cite{Baudis:2020nwe}. In addition, we apply corrections on the $S1$~and $S2$~signals as detailed in the same reference. Note that the SE~$S2$~signal is not subject to any drift time dependence. In this analysis, the PE"~scale of both $S1$~and $S2$~top signals is fixed utilising the single PE resolution of the SiPMs, see Fig.~\ref{fig:se_population}, right (cf.~Ref.~\cite{Baudis:2020nwe}, Sec.~5.1.2 for the $S1$~top correction). However, such a correction is not accessible for the bottom signals due to the lack of single PE resolution of the PMT.

\subsubsection{Electron extraction efficiency}
\label{subsubsec:electron_extraction_efficiency}

The electron extraction efficiency from the liquid to the gas phase can be assumed to be $\sim \SI{100}{\%}$ at a gas (liquid) extraction field of $\SI{10}{kV/cm}$ ($\SI{5.4}{kV/cm}$). We base this assumption on various former measurements that applied different methods. Among these are early absolute measurements directly comparing the ionisation signal below and above the liquid surface~\cite{Gushchin:1979,Aprile:2004_extraction}. In XENON100 a relative measurement was performed, comparing the recombination- and electron lifetime-corrected $S2$~signals of mono-energetic sources to the expected number of produced electrons~\cite{XENON100:2013wdu}. XENON1T compared the $g2$ obtained from the anti-correlation fit of several mono-energetic sources (assuming a literature $W$"~value) and the $g2$ from the lowest $S2$~signals~\cite{Aprile:2017aty}. However, more recent relative measurements imply that the extraction efficiency might only be $\SIrange{89}{95}{\%}$ at the extraction field of interest~\cite{Edwards:2017emx,Xu:2019dqb}. These were searching for saturation of the ratio of the $S2$~signals of mono-energetic calibration sources and the SE~$S2$ at increasing extraction fields. As noted in Ref.~\cite{Edwards:2017emx}, such measurements are, however, highly subject to systematic uncertainties from the geometry of the extraction region. In Ref.~\cite{Xu:2019dqb} the discrepancy to earlier relative measurements is attributed to scaling factors that arise when an independent determination of the number of initially produced electrons is lacking. In summary, the literature on this topic appears inconclusive and lacks a detailed unified explanation of the discrepancies between all the used methods. For this reason, we assume an electron extraction efficiency of $\SI{100}{\%}$ in this study and consider the consequence of this effect on our analysis. While a lower extraction efficiency would only diminish the statistics of the SE~population, it would reduce the charge yield of the calibration lines and thus, if disregarded, yield a higher $W$"~value when determined via Eqs.~\ref{eq:W_value_det1} and~\ref{eq:W_value_det2}. We further comment on the effect in Sec.~\ref{sec:discussion_conclusion}. 

\subsubsection{Liquid level}
\label{subsubsec:liquid_level}

The \ce{^{37}Ar} and \ce{^{83\text{m}}Kr} calibrations were conducted in two separate runs with a complete xenon recovery and filling procedure in between. The levelling procedure gives rise to an uncertainty of the liquid-gas interface of $\pm \SI{125} {\micro m}$ between the runs which is nominally kept in the middle of the gate and anode mesh. This in turn impacts the $S2$~signal amplification. A systematic study of this effect with \ce{^{83\text{m}}Kr} data shows that the levelling uncertainty corresponds to a maximum $S2$"~uncertainty of $\SI{2.5}{\%}$. However, we only include this uncertainty in the \ce{^{83\text{m}}Kr} data points of Fig.~\ref{fig:doke_plot} where we use information of both runs for the anti-correlation fit, and assume the liquid level to be constant within the runs, due to stable thermodynamic conditions, and uniform in the horizontal plane.

\begin{figure*}
\centering
\begin{subfigure}[b]{0.49\textwidth}
\includegraphics[width= \textwidth]{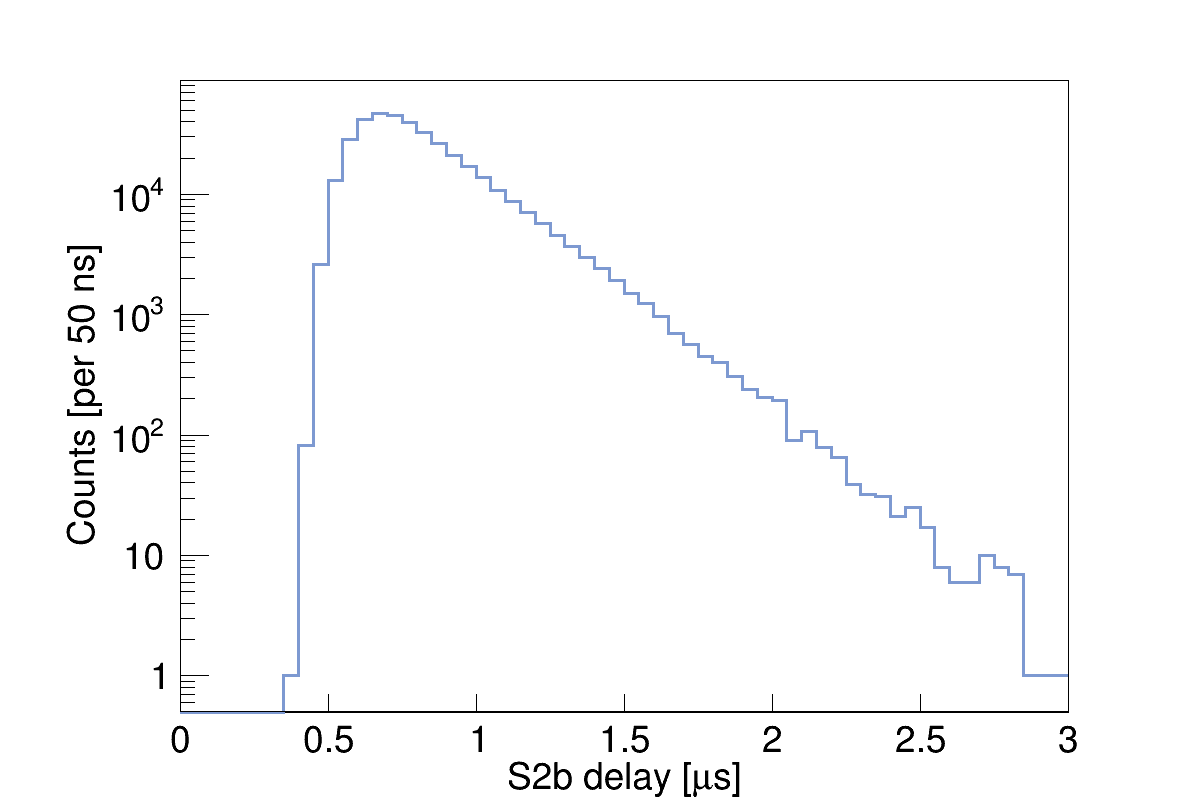}
\end{subfigure}
\begin{subfigure}[b]{0.49\textwidth}
\includegraphics[width= \textwidth]{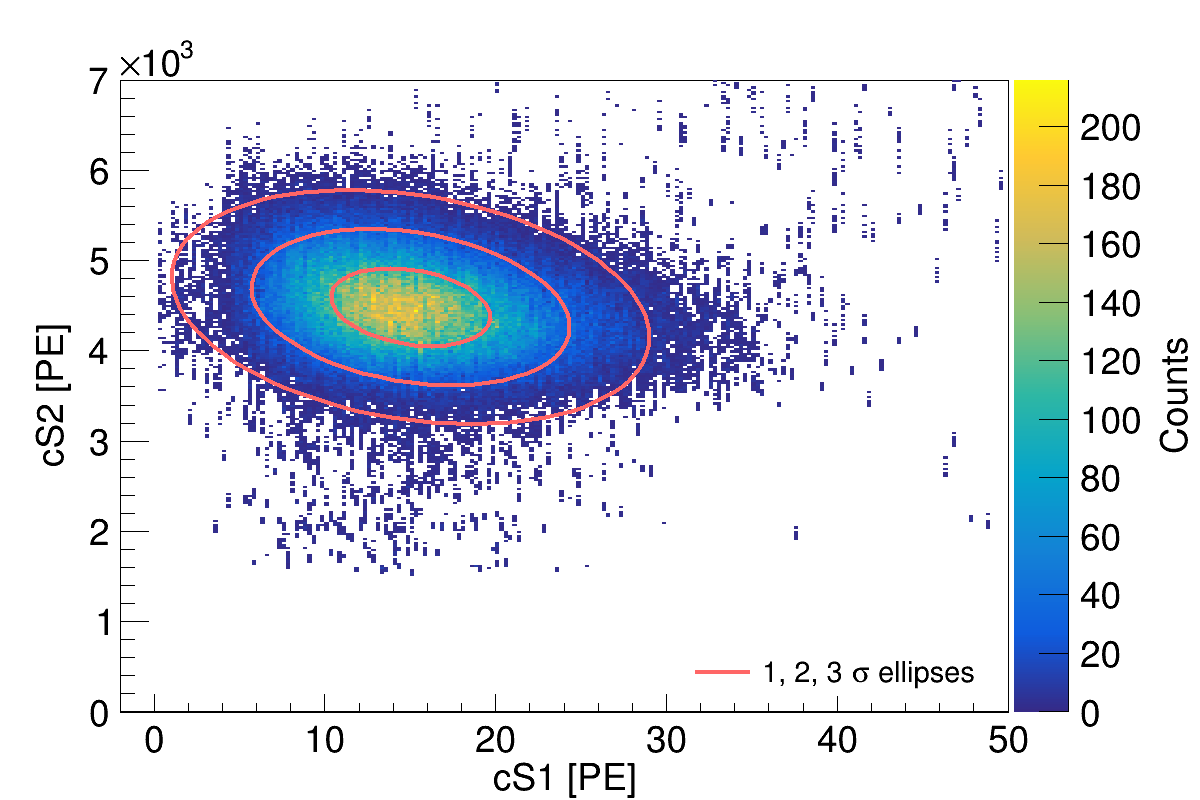}
\end{subfigure}
\caption{Left: \ce{^{83\text{m}}Kr} $S2$~bottom ($S2$b) delay histogram. Right: $\SI{2.82}{keV}$ event population of \ce{^{37}Ar} without DPE/crosstalk correction acquired at $\SI{968}{V/cm}$ drift field.}
\label{fig:s2_delay_K-shell_population}
\end{figure*}

\subsection{Photosensor effects}
\label{subsec:photosensor_effects}

We note that Eqs.~\ref{eq:W_value_det1},~\ref{eq:W_value_det2} are insensitive to an overall factor linear in $g1$, $g2$, $S1$, $S2$ that is constant in time, energy and common to the PMT and SiPMs. For instance, the result is insensitive to a scaling of the conversion from ADC bins to PE due to e.g.~the read-out electronics as long as it concerns both top and bottom sensors. However, as we discuss in Sec.~\ref{sec:discussion_conclusion}, the data shows a dependence of the relative light and charge yields of the top and bottom photosensors on the respective energy line, see Table~\ref{tab:yield_fractions}. Therefore, we do expect an influence of the hybrid photosensor configuration from systematically different characteristics of PMTs and SiPMs. Two classes of such characteristics can be distinguished: time-dependent and time-constant ones. Photosensor gains can generally show a time-dependence, e.g.~due to the aging of the photocathode~\cite{Aiello:2013wma}. Since we did not measure all input quantities for $W$ at the same point in time, a systematic gain change could impact the result even though all of the input quantities in the Eqs.~\ref{eq:W_value_det1} and~\ref{eq:W_value_det2} scale with the gain in the same fashion. All other characteristics treated subsequently are expected to be constant in time as long as the thermodynamic conditions are unchanged. For the considered interactions the photosensors are operated in the linear regime of their dynamical range and thus, we expect the characteristics to be independent of the interaction energy.

\subsubsection{Photosensor gain}
\label{subsubsec:photosensor_gain}

The photosensor gains were determined weekly with an LED calibration~\cite{Baudis:2020nwe} and have shown to be stable in time. We measured a PMT gain of $(3.76 \pm 0.06) \times 10^6$. The gain variations among the SiPM channels are small ($< \SI{6}{\%}$). The signals of the channels are however scaled individually with their mean gain. The error-weighted SiPM mean gain is $(3.12 \pm 0.01) \times 10^6$. The uncertainties on the gain measurements represent one standard deviation of the gain distributions of the photosensors which we use as an estimator for a systematic change over the period of data acquisition.

\subsubsection{Double photoelectron emission and crosstalk}
\label{subsubsec:DPE_crosstalk}

Double photoelectron emission (DPE), i.e.~the emission of two primary electrons originating from a single photon impacting the photocathode, is well-known for PMTs~\cite{Faham:2015kqa,LopezParedes:2018kzu,LUX:2019npm}. We use a DPE probability of $(20 \pm 5) \, \si{\%}$ for the R6041"~06~MOD PMT and correct for it. 
On the other hand, the working principle of SiPMs disfavours the existence of such an external enhancement effect. The output of a single cell is independent of the excess number of impacting photons and thus, light detection with SiPMs relies on the fact that, within their linear regime, the signal size is proportional to the number of triggered cells. The only known effect that increases the light signal of an isolated SiPM is the internal crosstalk among neighbouring cells, i.e.~from a photon crossing the trench and triggering another avalanche in a nearby cell. The internal crosstalk probability is usually determined from dark counts inside the cells which provides a good source of single PEs~\cite{Baudis:2018pdv}. We show in~\ref{app:SiPM_DPE} that for the S13371 SiPMs we do not see an excess with external scintillation light due to a DPE or DPE-mimicking effect beyond the internal crosstalk probability and conclude that the enhancement probability is $(2.2 \pm 0.1) \, \si{\%}$, an effect for which we correct as well.

Besides the internal crosstalk among neighbouring cells of a single sensor, SiPMs are known to feature an external crosstalk capability among different sensors~\cite{McLaughlin:2021xat}. Photons produced during electron avalanches can, instead of travelling to another cell, escape the sensor and eventually reach another one where they trigger a secondary avalanche. Such an effect must be estimated {\sl{in situ}} as it is geometry-dependent. To this end, we acquired scintillation-free dark count data with nitrogen gas under thermodynamic conditions very similar to those during the calibration runs using xenon~\cite{Baudis:2020nwe}. Since dark counts of different sensors are uncorrelated, the external crosstalk probability can be estimated from correlated events of a few PE in size. We obtain a mean external effective crosstalk probability of a SiPM in the top photosensor array of $(\SI{0.05}{} \pm \SI{0.01}{}) \, \si{\%}$. The light collection of the top SiPMs is thus expected to be enhanced by this amount. This result is corrected by the probability that a given sensor is in uncorrelated coincidence with any other sensor of the array within the examined trigger time window of $\SI{40}{ns}$. We measured this accidental probability to be $\SI{0.02}{\%}$ from coincidences outside of the trigger window, in agreement with expectation for a mean dark count rate of $(\SI{8.05}{} \pm \SI{0.03}{}) \, \si{Hz/mm^2}$~\cite{Baudis:2020nwe}. We do not expect the estimated external crosstalk probability among SiPM channels to change to a significant value, compared to the internal one, when the TPC is operated with xenon in dual-phase mode. During operation with LXe, an external SiPM crosstalk recorded by the PMT would be visible as a small coincident $S1$"~only signal in the PMT and a SiPM channel. However, we did not observe any event of this kind e.g.~during the \ce{^{37}Ar} calibration at $\SI{968}{V/cm}$ drift field. In summary, the external SiPM crosstalk effect can safely be neglected as it is subdominant to the internal one and no effect on the PMT is seen.

\subsubsection{Photon detection efficiency}
\label{subsubsec:PDE}

The manufacturer Hamamatsu Photonics claims for the VUV4 S13371 SiPMs a saturation photon detection efficiency (PDE) of $\SI{24}{\%}$ at $\SI{175}{nm}$ and $\SI{25}{\celsius}$ operated at $\SI{4}{V}$ overvoltage without crosstalk and afterpulsing~\cite{Hamamatsu_MPPC}. However, SiPM characterisations for the nEXO experiment imply a much lower saturation PDE of $\SIrange{9.9}{17.6}{\%}$, measured at $\SIrange{3.3}{3.8}{V}$ overvoltage and $\SI{233}{K}$, that differs by $\SI{8}{\%}$ among devices~\cite{Gallina:2019fxt,nEXO:2019jhg}. We are operating the SiPMs at more than $\SI{4}{V}$ overvoltage~\cite{Baudis:2020nwe}. In view of these measurements, we can thus assume a typical PDE of $\SI{18}{\%}$. Hamamatsu Photonics provided us with the spectral response curve of the R6041"~06~MOD PMT that shows a peak quantum efficiency (QE) of $\sim \SI{30}{\%}$ at $\SI{175}{nm}$. The MOD~specification sets a minimum QE of $\SI{25}{\%}$ at that wavelength as requirement. An independent measurement shows a QE of $\sim \SI{28}{\%}$~\cite{Arazi:2013fxa}. The electron collection efficiency of that PMT was specified with $\sim \SI{70}{\%}$ upon enquiry to the manufacturer. We can thus conclude that the typical PDE of that PMT is $\SI{20}{\%}$. Since the SiPMs and the PMT have similar typical PDEs with a percent-level difference, systematics from the PDE difference are negligible for this analysis.  

\subsubsection{Infrared sensitivity}
\label{subsubsec:infrared_sensitivity}

Apart from emitting light in the VUV~region, GXe is known to emit significantly in the infrared (IR)~\cite{Carugno:1998tb,BRESSI2000254}: it features a strong line at $\SI{1300}{nm}$ wavelength at a VUV-comparable zero-field light yield of $(21 \pm 3)\,\si{photons / keV}$\footnote{This is only a lower limit measured at $\SIrange{700}{1600}{nm}$ but can be assumed to be the true value considering Ref.~\cite{Bressi:2001st}.} at $\SI{2}{bar}$, measured with $\alpha$-particles at $\SI{4.3}{MeV}$~\cite{BELOGUROV2000167}. LXe features a different IR~scintillation spectrum which emits mostly below $\SI{1200}{nm}$ but with a very poor yield~\cite{Bressi:2001st}. For this reason, only the $S2$~light can contain a significant amount of IR~radiation. However, both photosensor types are insensitive at these wavelengths and thus IR-scintillation does not contribute significantly to our signals. For the SiPMs, the band gap of silicon sets a sensitivity cutoff at $\sim \SI{1100}{nm}$. According to the manufacturer, the PMT is insensitive beyond $\SI{1000}{nm}$ which is confirmed by the trend of the spectral response curve in Ref.~\cite{Schrott:2021flv} that shows a vanishing QE for wavelengths above $\SI{650}{nm}$.  

\subsection{Data acquisition and processing effects}
\label{subsec:processing_effects}

\subsubsection{Detection and tagging efficiency}
\label{subsubsec:detection_tagging_efficiency}

As mentioned in Sec.~\ref{subsec:data}, the DAQ settings during the \ce{^{37}Ar} calibration period allowed us to trigger on $\SI{7}{mV}$ PMT signals. Naturally, the trigger efficiency of signals that have comparable or lower amplitudes is reduced which is particularly relevant for SE~$S2$~signals. The translation from signal amplitude to signal charge depends on the signal type. We employ a data-driven approach for the SE~population, using their amplitude-over-width ratio as discriminator to obtain a detection efficiency, as shown in Fig.~\ref{fig:se_population}, left. To this end, we consider the fraction of SE~peaks that did not issue a trigger in events in which a subsequent larger signal triggered. Based on the same discriminator, we obtain the tagging efficiency of the processor to correctly identify SE~events as $S2$~signals, see the same figure. We correct the SE~population for these efficiencies and assume that these are $\SI{100}{\%}$ for larger signals.

\subsubsection{Peak-splitting routine}
\label{subsubsec:peak-splitting_routine}

In order to separate the distinct peaks of the $\SI{32.15} {keV}$ and the $\SI{9.41} {keV}$ line of \ce{^{83\text{m}}Kr}, our raw-data processor splits at the intermediate minimum whenever a moving average of the waveform has fallen below half of maximum values of both peaks. This algorithm was chosen for its simplicity and good performance as it avoids fitting and has no need for a waveform template. Assuming Gaussian shape, we estimated in Ref.~\cite{Baudis:2020nwe} the typical leakage into the neighbouring peak, caused by the processor, to be $\leq \SI{10}{\%}$.   

Here, we use a data-driven approach based on the actual shape of the photosensor signals to obtain a reliable estimate of the maximally possible leakage between the two peaks. To this end, we select a few hundred well-separated $S2$~signal pairs with a delay of at least $\SI{1.2} {\micro s}$\footnote{The narrow \ce{^{83\text{m}}Kr} $S2$"~width distribution is maximal at $\SI{0.6} {\micro s}$ and thus for delays $> \SI{1.2} {\micro s}$, the waveform has basically fallen to baseline-level in between the peaks.} and shift the small $\SI{9.41}{keV}$ $S2$ towards the large $\SI{32.15}{keV}$ $S2$ up to the point where splitting, according to the algorithm, is still just about possible. Integration of the charge and comparison to the isolated peaks confirms the expectation that the leakage is directed from the large to the small $S2$~signal. The maximum errors of the PMT (SiPM) signals are $(-2.0 \pm 1.3) \,\%$ $\left( (-3.7 \pm 1.9) \, \% \right)$ for the $\SI{32.15}{keV}$ line and $(+10.2 \pm 5.9) \, \%$ $\left( (+19.2 \pm 8.4) \% \right)$ for the $\SI{9.41}{keV}$ line. A positive (negative) error indicates here that the $S2$~peak is reconstructed with a higher (lower) charge. Since a correction for any given $S2$"~delay would be rather uncertain, we will propagate these maximum errors. 

We observe that the splitting algorithm only allows for a minimal \ce{^{83\text{m}}Kr} $S2$"~delay of $\sim \SI{400}{ns}$ (see Fig.~\ref{fig:s2_delay_K-shell_population}, left) at which $S1$~signals, that feature widths in the range $\SIrange{60}{140}{ns}$, are well-separated. The delay distribution peaks at $\sim \SI{700}{ns}$ and falls exponentially towards higher separation. At such large separations, no delay dependence of the $S1$~signals could be observed, unlike at shorter delays where corrections are needed~\cite{Baudis:2013cca}.

\subsubsection{Fitting procedure}
\label{subsubsec:fitting_procedure}

The two-dimensional $\SI{2.82}{keV}$ distribution in $S1$"~$S2$"~space, shown in Fig.~\ref{fig:s2_delay_K-shell_population}, right, features a slight asymmetry (skewness) that was also observed elsewhere~\cite{Boulton:2017hub,Szydagis:2021hfh}. This is due to DAQ and processing efficiencies, the low light level or charge loss and can be modelled with skew-Gaussians~\cite{LUX:2020car}. However, here the shape of the distribution is of less relevance to the analysis for we only extract its mean. It is thus sufficient to use a standard Gaussian fit and instead estimate the uncertainty on the mean. To this end, and in addition to the standard fit parameter errors, we systematically varied the fit interval, including asymmetric intervals. This effect is more pronounced in the $S1$- than in the $S2$"~direction and accounts for an $S1$"~uncertainty of $^{+0.7}_{-4.1} \, \si{\%}$ and an $S2$"~uncertainty of $^{+0.2}_{-0.7} \, \si{\%}$. In Ref.~\cite{Szydagis:2021hfh} the larger $S1$~bias is explained by the fact that only upward $S1$~fluctuations above the $S1$~threshold are measured at such low light levels.

\begin{table}
\centering
\begin{tabularx}{\columnwidth}{lX}
\hline\noalign{\smallskip}
\bf{Systematic effect}  & \bf{Relevance and treatment}\\
\noalign{\smallskip}\hline\noalign{\smallskip}
\bf{TPC effects}        & \\
LXe purity (electron lifetime) & Drift time dependence of top and bottom $S2$~signals corrected \\
TPC geometry & Drift time dependence of bottom $S1$~signals corrected, active volume fiducialised in radius ($< \SI{10}{mm}$) and depth ($-[29,2] \, \si{mm}$) \\
Electron extraction efficiency &  Assumed to be $\SI{100}{\%}$\\
Liquid level & Assumed constant within runs, maximum uncertainty of $\SI{2.5}{\%}$ included in \ce{^{83\text{m}}Kr} data in Fig.~\ref{fig:doke_plot} \\
\noalign{\smallskip}\hline\noalign{\smallskip}
\bf{Photosensor effects}    & \\
Photosensor gain & Channels individually scaled for their mean gain, $1 \sigma$ of gain distribution assumed as variation in time and included in result \\
DPE and crosstalk & Result corrected for PMT DPE of $(20 \pm 5) \, \si{\%}$ and internal SiPM crosstalk of $(2.2 \pm 0.1) \, \si{\%}$, external SiPM crosstalk of $(\SI{0.05}{} \pm \SI{0.01}{}) \, \si{\%}$ neglected\\
PDE & PMT/SiPM difference negligible \\
IR sensitivity & Negligible \\
\noalign{\smallskip}\hline\noalign{\smallskip}
\bf{DAQ and processing effects}     & \\
Detection/tagging efficiency & Corrected for SE, see Fig.\ref{fig:se_population}, left, assumed to be $\SI{100}{\%}$ at higher energies \\
Peak-splitting routine & Relevant for \ce{^{83\text{m}}Kr} $S2$~signals, maximum uncertainty for PMT (SiPM) included in result: $(-2.0 \pm 1.3) \,\%$ $\left( (-3.7 \pm 1.9) \, \% \right)$ for $\SI{32.15}{keV}$, $(+10.2 \pm 5.9) \, \%$ $\left( (+19.2 \pm 8.4) \% \right)$ for $\SI{9.41}{keV}$ \\
\ce{^{83\text{m}}Kr} $S1$"~delay dependence & Not observed  \\
Fitting procedure & Parameter uncertainties included; for asymmetric $\SI{2.82}{keV}$ distribution, maximum uncertainty on $S1$ ($S2$) mean from systematic variation of fit intervals included in result: $^{+0.7}_{-4.1} \, \si{\%}$ ($^{+0.2}_{-0.7} \, \si{\%}$)\\
\noalign{\smallskip}\hline
\end{tabularx}
\caption{Summary of the considered systematic effects and applied corrections.}
\label{tab:systematics}
\end{table}
\section{Result}
\label{sec:result}

We evaluate Eq.~\ref{eq:W_value_det2} for the measurements of Sec.~\ref{sec:measurements}, considering the systematics and corrections of the previous section, and obtain $W=\SI{11.5}{} \, ^{+0.2}_{-0.3} \, \mathrm{(syst.)} \, \si{\electronvolt}$. We do not quote a statistical uncertainty as it is subdominant compared to the systematic effects which are mainly due to the photosensor gain uncertainties, and the parameter errors of the anti-correlation fit. The latter come mostly from the liquid level uncertainty and the \ce{^{83\text{m}}Kr}"~$S2$~splitting routine. The overall uncertainty is obtained from a propagation of the contributing uncertainties. The correlations of the inputs are maximal, however, these do not increase the final uncertainty much. Determining the $W$"~value with the local approach using Eq.~\ref{eq:W_value_det1} at any of the considered $S1$"~$S2$"~space populations of the calibration sources yields compatible results with mean values in the range $W=\SIrange{11.1}{11.6}{\electronvolt}$ with slightly higher errors due to the less direct nature of the approach.
\section{Discussion and conclusion}
\label{sec:discussion_conclusion}

We observe slightly different relative light and charge yields of the top and bottom photosensors, depending on the selected event population, see Table~\ref{tab:yield_fractions}. As there is no fundamental physical reason for such differences, we attribute these to systematic effects discussed in Sec.~\ref{sec:systematics_corrections}. For instance, the peak-splitting routine increases the charge yield of the $\SI{9.41}{keV}$ \ce{^{83\text{m}}Kr} line while it slightly reduces the one of the $\SI{32.15}{keV}$ line and leaves the light yield of both lines unchanged (see Sec.~\ref{subsubsec:peak-splitting_routine}). The slightly higher relative PMT charge yield in SE~events could be due to an unquantified systematic deficiency of our processing framework, leading to a higher (smaller) reconstructed charge in the single PE regime of the bottom (top) sensors. However, in lack of a thorough explanation of the deviations shown in Table~\ref{tab:yield_fractions}, we would like to remark that a physical effect enhancing (reducing) the photosensor charge yield in the SE~regime would increase (decrease) $g2$ and the resulting W"~value. All photosensor effects were weighted with the relative yields in Table~\ref{tab:yield_fractions} when applied to the combined, top and bottom, signals. We note however, that the influence of the hybrid photosensor characteristics on the final result is suppressed by the fact that the weights are very similar and the effect would in particular vanish for equal weights, cf.~discussion in Sec.~\ref{subsec:photosensor_effects}. 

As mentioned in Sec.~\ref{subsubsec:electron_extraction_efficiency}, we did not consider an electron extraction deficiency in this study. We note however, that the resulting $W$"~value would only be lower than the obtained one when correcting for an imperfect extraction efficiency. Because the incorporation of this systematic is a straightforward scaling of the charge yields of the \ce{^{37}Ar} and \ce{^{83\text{m}}Kr} lines and thus, of the $W$"~value, we do not include the effect of a potential non-unity extraction efficiency in the lower error bound of the final result.

\begin{table}
\centering
\begin{tabular}{ccc}
\hline\noalign{\smallskip}
\bf{Energy}     & \bf{LY fraction [\%]}    & \bf{QY fraction [\%]}\\
\bf{[keV]}      & \bf{SiPMs / PMT}           & \bf{SiPMs / PMT} \\
\noalign{\smallskip}\hline\noalign{\smallskip}
SE      & -- / -- & 18--19 / 81--82 \\
2.82    & 10--11 / 89--90  & 25--27 / 73--75 \\ 
9.41    & 7 / 93           & 20--21 / 79--80 \\
32.15   & 7--8 / 92--93   & 31--32 / 68--69 \\
41.56   & 7--8 / 92--93   & 29--30 / 70--71 \\
\noalign{\smallskip}\hline
\end{tabular}
\caption{Fractions of the top SiPMs / bottom PMT sensors on the total light (LY) and charge yields (QY) for SE~events, the $\SI{2.82}{keV}$ line of \ce{^{37}Ar} as well as the split $\SI{32.15}{keV}$ and $\SI{9.41}{keV}$ and the merged $\SI{41.56}{keV}$ line of \ce{^{83\text{m}}Kr}. The yields are independent of the drift field and the small variations among the measurements are represented by the displayed ranges.}
\label{tab:yield_fractions}
\end{table}

Based on SE~events and data from low-energy internal \ce{^{37}Ar} and \ce{^{83\text{m}}Kr} calibration sources in a small dual-phase TPC, we have measured the mean electronic excitation energy of LXe to be $W=\SI{11.5}{} \, ^{+0.2}_{-0.3} \, \mathrm{(syst.)} \, \si{\electronvolt}$, with negligible statistical uncertainty. Although Eqs.~\ref{eq:W_value_det1} and~\ref{eq:W_value_det2} are widely insensitive to the different photosensor characteristics of PMT and SiPMs, we carefully treated the known systematic deviations. In particular, we presented in~\ref{app:SiPM_DPE} a combinatorial approach to search for a DPE effect in SiPMs based on SE~data and found no excess beyond the known crosstalk effect. Additionally, we considered the systematic uncertainties from TPC, DAQ and data processing effects. 

Albeit affected by various sources of systematics, our result features a competitive uncertainty range compared to former measurements. It is compatible with the $W$"~value of $(11.5 \pm 0.1\,\mathrm{(stat.)} \pm 0.5\, \mathrm{(syst.)}) \, \si{eV}$ reported by the EXO"~200 Collaboration~\cite{EXO-200:2019bbx} at $\mathcal{O}(\SI{1}{MeV})$ energies which, thus, is reproducible at keV-scale energies. The EXO"~200 experiment was a single-phase LXe detector with a wire charge readout. An external charge-injection allowed for an absolute calibration of the amplifier on the readout plane. The uncertainty on the EXO"~200 result is dominated by the calibrations of the light and charge response of the detector.  

However, the $W$"~value found here is lower than determined in other former measurements~\cite{Doke:2002,Shutt:2006ed}. In particular, it is incompatible with the established value measured by E.~Dahl of $(\SI{13.7}{} \pm \SI{0.2}{}) \, \si{eV}$~\cite{Dahl:2009nta} which was determined in a LXe detector of comparable size with $\SI{122}{keV}$ and $\SI{136}{keV}$ gamma rays from an external \ce{^{57}Co} source. The detector was operated in dual-phase and single-phase mode. For the latter, an amplifier on the anode allowed for a direct calibration of the charge yield used to measure the gain $g2$. The calibration of the amplifier was also the dominant source of uncertainty for the final result. In dual-phase mode, the detector was operated at the same extraction field as Xurich~II and thus, the higher value found by E.~Dahl cannot be explained by a lower electron extraction efficiency. The difference to our obtained value would stay constant for lower extraction efficiencies. 

In order to check whether this discrepancy to former measurements is due to a higher observed absolute yield in the scintillation or ionisation channel, we can compare our result to predictions of the Noble Element Simulation Technique (NEST)~\cite{Szydagis:2011tk}. From our measurements of $g2$ and $g2/g1$ in Sec.~\ref{sec:measurements} we can calculate the corrected scintillation gain to be $g1=\SI{0.103}{} \, ^{+0.003}_{-0.004} \, \si{PE/\gamma}$. Using the definitions of $g1$ and $g2$ (cf.~Sec.~\ref{subsec:method}), we obtain at $\SI{2.82}{keV}$ and $\SI{968}{V/cm}$ drift field on average the following number of excitation quanta from our measurements: $n_{\gamma}= 115.6$ and $n_{\mathrm{e}^{-}}=130.4$. At this energy and drift field, and for the LXe density given in Sec.~\ref{subsec:setup}, the NEST calculator~\cite{NEST:Calculator} predicts absolute yields of $n_{\gamma}^{\mathrm{NEST}}= 85.8 \, (71.1)$ and $n_{\mathrm{e}^{-}}^{\mathrm{NEST}}=120.2 \, (134.9)$, respectively, for the gamma (beta) model. Thus, the discrepancy is mostly seen in the scintillation channel and we observed $S1$~signals with a higher yield than predicted by NEST. Qualitatively, the same is seen at higher energies or lower drift fields.
 
A change of the numerical value of $W$ does not influence the translation from $S1$~and $S2$~signals into deposited energy when performing measurements relative to calibration lines. According to Eq.~\ref{eq:W_value_det1}, a lower $W$"~value will generally reduce the microscopic detector response parameters~$g1$ and~$g2$ and thus, the absolute energy response to excitation quanta. As we have seen from the comparison with NEST, the rescaling will be mostly in the $g1$~value. 

The Fano factor $F$ accounts for the non-Poissonian nature of the intrinsic fluctuation of the number of quanta $n \coloneqq n_{\gamma}+n_{\mathrm{e}^{-}}$ produced in an interaction~\cite{Fano:1947zz,Doke:1976zz,Carmo:2008}. At the recombination-independent combined scintillation and ionisation energy scale of ER interactions (cf.~Eq.~\ref{eq:W_value_def}), we can write:
\begin{equation}
\sigma_{n}= \sqrt{Fn} \quad.
\end{equation}
This expression is related to the (intrinsic) energy resolution of an ideal detector at energy $E$ (Fano limit~\cite{Aprile:2009dv}):
\begin{equation}
\frac{\sigma_{E}}{E}=\frac{\sigma_{n}}{n} \quad.
\end{equation}
Using Eq.~\ref{eq:W_value_def}, we obtain 
\begin{equation}
\sigma_{E}=\sqrt{FEW} \quad. 
\end{equation}
Thus, for a measured energy resolution, the Fano factor can be determined based on $W$~\cite{Seguinot:1995uf}, and a lower value of $W$ would imply a higher Fano factor. 

Due to these implications, we expect our work to provoke other independent measurements of the $W$"~value and studies as to the origin of the deviation of almost $\SI{20}{\%}$ compared to other low-energy measurements. New measurements of the $W$"~value in LXe would also provide valuable input to NEST~\cite{Szydagis:2011tk}, which aims to model the absolute light and charge yields in LXe detectors for different energies and electric drift fields.

\begin{acknowledgements}
This work was supported by the European Research Council (ERC) under the European Union's Horizon 2020 research and innovation programme, grant agreement No. 742789 ({\sl Xenoscope}). We thank Nastassia Grimm for her help with the combinatorics for the SiPM DPE analysis and Marc Schumann for helpful discussions. We also thank Rafael F.~Lang and Abigail Kopec for suggesting a study as to an external DPE effect of SiPMs as a potential systematic effect. 
\end{acknowledgements}

\appendix
\section{Search for DPE in SiPMs}
\label{app:SiPM_DPE}

We apply a combinatorial approach on the SE~population to determine the probability $q$ that a $\SI{2}{PE}$ signal originates from one initial incoming photon onto the photocathode. We assume that $q$ is universal to all channels. Given a fiducial radius, let $\{p_{i}\}_{i=0,\ldots,15}$ be the set of mean light fractions of the 16~SiPMs for homogeneously distributed SE~events. Furthermore, we denote the number of recorded hits in sensor $i$ by $k_{i}$. For good statistics in the SE~population, we consider events with 3~detected hits and we call the total number of these events $N^{3\mathrm{h}}$. Then, we can distinguish three cases\footnote{A two-channel coincidence requirement for event building is based on narrow coincidence time windows. This reduces the contribution of randomly-coinciding dark counts to a negligible level~\cite{Baudis:2020nwe}.}:

\paragraph{I. \hspace{0.2 cm} $k_{i}=3$, $k_{j}=0 \quad \forall j \neq i$.}

This signal can be produced by 3, 2~or 1~initial photons. The latter process is next-to-next-to-leading order and thus highly suppressed. However, we will discard this first case as no such events were recorded, for it is a highly improbable event topology, and it is thus not accessible here. Note, that this is not due to the twofold coincidence requirement of the event building because additional PMT-triggers are allowed.   

\paragraph{II. \hspace{0.2 cm} $k_{i}=2$, $k_{l}=1$, $k_{j}=0 \quad \forall j \neq l \neq i$.}

This signal can be produced by 3~or 2~initial photons. We expect the following number of events with 2~hits in one and 1~hit in another sensor:
\begin{equation}
\label{eq:II_case}
N^{\mathrm{II}}=
\begin{cases}
N^{3\mathrm{h}} \cdot 3\sum \limits_{i=0}^{15} p_{i}^2(1-p_{i}), \ q=0 \\
N_{3 \, \gamma} \cdot 3\sum \limits_{i=0}^{15} p_{i}^2(1-p_{i})+N_{2 \, \gamma}^{2\mathrm{s}} \cdot 2q(1-q), \ q>0
\end{cases} \quad.
\end{equation}
Here, $N_{3 \, \gamma}$ denotes the number of events with 3~initial photons that are all detected and $N_{2 \, \gamma}^{2\mathrm{s}}$ is the number of events with 2~initial photons that are both detected by two different sensors. If $N^{2\mathrm{h}, \, 2\mathrm{s}}$ is the number of events with total 2~hits in 2~sensors, i.e.~one in each, we can identify
\begin{equation}
\label{eq:N2}
N_{2 \, \gamma}^{2\mathrm{s}}=\frac{N^{2\mathrm{h}, \, 2\mathrm{s}}}{(1-q)^2} \quad.
\end{equation}

\paragraph{III. \hspace{0.2 cm} $k_{i}=k_{l}=k_{m}=1$, $k_{j}=0 \quad \forall j \neq l \neq m \neq i$.}

This signal can only be produced by 3~initial photons. We expect the following number of events with 3~hits in three sensors, i.e.~one in each:
\begin{equation}
\label{eq:III_case}
N^{\mathrm{III}}=
\begin{cases}
N^{3\mathrm{h}} \cdot \sum \limits_{i=0}^{15} \sum \limits_{j \neq i} p_{i}p_{j}(1-p_{i}-p_{j}), \ q=0 \\
N_{3 \, \gamma} \cdot \sum \limits_{i=0}^{15} \sum \limits_{j \neq i} p_{i}p_{j}(1-p_{i}-p_{j}), \ q>0
\end{cases} \quad.
\end{equation}

Combining the Eqs.~\ref{eq:II_case},~\ref{eq:N2} and~\ref{eq:III_case} for $q>0$, we obtain
\begin{equation}
q=\frac{\tilde N}{2N^{2\mathrm{h}, \, 2\mathrm{s}}+\tilde N} \quad,
\end{equation}
\begin{equation}
\tilde N \coloneqq N^{\mathrm{II}}-N^{\mathrm{III}}\frac{3\sum \limits_{i=0}^{15} p_{i}^2(1-p_{i})}{\sum \limits_{i=0}^{15} \sum \limits_{j \neq i} p_{i}p_{j}(1-p_{i}-p_{j})} \quad.
\end{equation}
The following mean light collection fractions of the sensors, that are arranged in a $4 \times 4$ array, can be deduced from SE~data for the same fiducial volume as used in the analysis above: $\SI{4.6}{\%}$ (corner sensors), $\SI{6.2}{\%}$ (edge sensors), $\SI{8.0}{\%}$ (middle sensors). Those match well light simulations with the GEANT4 particle physics simulation kit~\cite{GEANT4:2002zbu}. Integrating the SE~spectra with the respective data selection cuts (similar to Fig.~\ref{fig:se_population}, right) for $N^{\mathrm{II}}$, $N^{\mathrm{III}}$ and $N^{2\mathrm{h}, \, 2\mathrm{s}}$ yields $q=(2.2 \pm 0.1) \, \si{\%}$, with Poissonian uncertainty. The crosstalk probability of the VUV4 S13371 SiPMs was measured to be $\SI{2.1}{\%}$ at $\SI{4}{V}$ overvoltage and $\SI{3.3}{\%}$ at $\SI{5}{V}$ overvoltage, and is widely constant with temperature~\cite{Baudis:2018pdv}. We operated the SiPMs in Xurich~II at $\SI{51.5}{V}$ which corresponds to $\SI{5}{V}$ overvoltage at a temperature of $\SI{190}{K}$ measured in the gas phase. However, due to the heat dissipation of the pre-amplifiers of $\sim \SI{3}{W}$, the temperature at the sensors is expected to be slightly higher~\cite{Baudis:2020nwe}. In fact, the overvoltage is reduced by $\SI{1}{V}$ below $\SI{220}{K}$. In this regard, the determined probability $q$ is well within the range of the expected crosstalk and we do not see an excess that points towards an external SiPM DPE effect. Moreover, we can conclude that the crosstalk determination based on dark counts inside the cells is well-motivated as the result does not change when using an external source of single scintillation photons.

\bibliographystyle{JHEP}
\bibliography{W-value}  

\end{document}